\documentclass[]{interact}

\usepackage{epstopdf}
\usepackage[caption=false]{subfig}

\usepackage[numbers,sort&compress]{natbib}
\bibpunct[, ]{[}{]}{,}{n}{,}{,}
\renewcommand\bibfont{\fontsize{10}{12}\selectfont}
\makeatletter
\def\NAT@def@citea{\def\@citea{\NAT@separator}}
\makeatother

\theoremstyle{plain}

\theoremstyle{definition}

\theoremstyle{remark}

\begin{document}


\title{Temporal analysis of radiating current densities}


\author{
\name{Wei Guo\textsuperscript{a}\thanks{CONTACT W. Guo. Email: weiguoguo@hotmail.com} }
\affil{\textsuperscript{a}P. O. Box 470011, Charlotte, North Carolina 28247, USA}
}

\maketitle

\begin{abstract}
From electromagnetic wave equations, it is first found that, mathematically, any current density that emits an electromagnetic wave into the far-field region has to be differentiable in time infinitely, and that while the odd-order time derivatives of the current density are built in the emitted electric field, the even-order derivatives are built in the emitted magnetic field.  With the help of Faraday's law and Amp\`{e}re's law, light propagation is then explained as a process involving alternate creation of electric and magnetic fields.  From this explanation, the preceding mathematical result is demonstrated to be physically sound.  It is also explained why the conventional retarded solutions to the wave equations fail to describe the emitted fields.
\end{abstract}



In electrodynamics \cite{pan62,jac75,zan12}, a time-dependent current density $\vec{j}(\vec{r},t)$ and a time-dependent charge density $\rho (\vec{r},t)$, all evaluated at position $\vec{r}$ and time $t$, are known to be the sources of an emitted electric field $\vec{E}$ and an emitted magnetic field $\vec{B}$: 
\begin{equation}
\label{e1}
\nabla ^{2}\vec{E}-\frac{1}{c^{2}}\frac{\partial ^{2}}{\partial t^{2}}\vec{E}=\frac{4\pi}{c^{2}}\frac{\partial }{\partial t}\vec{j}+4\pi \nabla \rho,    
\end{equation}
and 
\begin{equation}
\label{e2}
\nabla ^{2}\vec{B}-\frac{1}{c^{2}}\frac{\partial ^{2}}{\partial t^{2}}\vec{B}=-\frac{4\pi}{c^{2}}\nabla \times \vec{j},
\end{equation}
where $c$ is the speed of these emitted fields in vacuum.  See Refs. \cite{zan12,roh02} for derivation of these equations.  (In some theories \cite{chu96}, on the other hand, $\rho$ and $\vec{j}$ are argued to be responsible for instantaneous action-at-a-distance fields, not for fields propagating with speed $c$.) Note that when the fields are observed in the far-field region, the contribution to $\vec{E}$ from $\rho$ can be practically ignored \cite{jef04}, meaning that, in that region, Eq. (\ref{e1}) simplifies to
\begin{equation}
\label{e3}
\nabla ^{2}\vec{E}-\frac{1}{c^{2}}\frac{\partial ^{2}}{\partial t^{2}}\vec{E}=\frac{4\pi}{c^{2}}\frac{\partial }{\partial t}\vec{j},
\end{equation}
and the current density becomes the only source of electromagnetic radiation.  In the following, the discussion is limited in the far-field region for simplicity.  See, nevertheless, Ref. \cite{nov07} for a review of near-field optics.

In $\vec{j}(\vec{r},t)$, the argument $\vec{r}$ defines a limited region in which the current density is confined.  The other argument $t$, on the other hand, determines how the current density changes with time.  Unlike $\vec{r}$, however, $t$ can in principle take any value.  Then one question arises.  Does this observation necessarily mean that there are no restrictions on the time evolution of a radiating current density?  A glance at Eq. (\ref{e3}) immediately shows that there are restrictions on the current density---the current density has to have the first-order time derivative to the minimum.  Thus, not every time-dependent current density can serve as a source of electromagnetic radiation.

Still in Eq. (\ref{e3}), since the emitted electric field $\vec{E}$ must be a function of $\partial \vec{j}/\partial t$ on the one hand, and has to be differentiable twice in time on the other hand, the current density $\vec{j}$ must, in addition, have the third-order time derivative.  Similarly, from Eq. (\ref{e2}), the wave equation satisfied by the emitted magnetic field, it has to be concluded that the current density must be differentiable in time twice.  These straightforward observations can be extended to two hypotheses.  First, a radiating current density should have high-order time derivatives.  Second, while the odd-order time derivatives of the current density are associated with $\vec{E}$, the even-order derivatives are associated with $\vec{B}$.  In the following, these hypotheses are examined mathematically and physically.

When the second term on the left-hand side of Eq. (\ref{e3}) is moved to the right-hand side, the wave equation formally becomes a Poisson equation:
\begin{equation}
\label{e4}
\nabla ^{2} \vec{E}=\frac{1}{c^{2}}\frac{\partial ^{2}}{\partial t^{2}}\vec{E}+\frac{4\pi}{c^{2}}\frac{\partial }{\partial t}\vec{j}.
\end{equation}
Then, with the help of the following relation \cite{jac75}, which defines the Green's function of the operator $\nabla ^{2}$ \cite{arf85,jen18},
\begin{equation}
\label{e5}
\nabla ^{2}\frac{1}{\vert \vec{r}-\vec{r}_{1}\vert}=-4\pi \delta (\vec{r}-\vec{r}_{1}),    
\end{equation}
it is convenient to find one particular solution of $\vec{E}$ from Eq. (\ref{e4}):
\begin{eqnarray}
\label{e6}
\vec{E}(\vec{r},t)&=&-\frac{1}{4\pi c^{2}}\int _{V_{1}} \frac{1}{\vert \vec{r}-\vec{r}_{1}\vert }\frac{\partial ^{2}}{\partial t^{2}}\vec{E}(\vec{r}_{1},t)d\vec{r}_{1}\nonumber\\
& &-\frac{1}{c^{2}}\int _{V_{2}}\frac{1}{\vert \vec{r}-\vec{r}_{1}\vert }
\frac{\partial }{\partial t}\vec{j}(\vec{r}_{1},t)d\vec{r}_{1},  \end{eqnarray}
where while $V_{2}$ denotes the volume occupied by the current density, $V_{1}$ is some simply connected volume enclosing $\vec{r}$.  The solution in Eq. (\ref{e6}) is in fact an integral equation of $\vec{E}$.  In the literature, different integral equations are constructed for $\vec{E}$.  In Ref. \cite{zan12}, for example, the integral equation contains not only volume integrals but also surface integrals, and initial conditions have to be employed to remove the latter integrals.   

Through iteration, $\vec{E}$ in Eq. (\ref{e6}) is expanded into a series, from which it turns out that $\vec{E}$ depends not just on the first-order and third-order time derivatives of the current density---$\vec{E}$ depends on all odd-order time derivatives of the current density:
\begin{eqnarray}
\label{e7}
\vec{E}(\vec{r},t)&=&-\frac{1}{c^{2}}\int _{V_{2}}\frac{1}{\vert \vec{r}-\vec{r}_{1}\vert }\frac{\partial }{\partial t}\vec{j}(\vec{r}_{1},t)d\vec{r}_{1}\nonumber\\
&&+\frac{1}{4\pi c^{4}}\int _{V_{1}}\frac{1}{\vert \vec{r}-\vec{r}_{1}\vert }d\vec{r}_{1}\int _{V_{2}}\frac{1}{\vert \vec{r}_{1}-\vec{r}_{2}\vert }\nonumber\\
& &\times \frac{\partial ^{3}}{\partial t^{3}}
\vec{j}(\vec{r}_{2},t)d\vec{r}_{2}+\cdots .
\end{eqnarray}
It is a straightforward matter to confirm that the preceding serial solution does solve the electric wave equation in Eq. (\ref{e3}).  Of course, the serial solution has to be convergent, and the convergence is guaranteed when $V_{1}$ is sufficiently small.  Similarly, it is found from Eq. (\ref{e2}) that $\vec{B}$ depends on all even-order time derivatives of $\vec{j}$.  Thus, at least from the viewpoint of mathematics, in order for the electric and magnetic fields to be emitted out of the current density into the far-filed region in the form of a wave, the current density has to be differentiable in time infinitely.  Still from the viewpoint of mathematics, while the emitted electric field depends on the odd-order time derivatives of the current density, the emitted magnetic field depends on the even-order derivatives.  The hypotheses made early in the present work are well founded in mathematics.

It is natural to ask if the hypotheses also have any significance in physics.  From the viewpoint of physics, the wave equations, on which the preceding mathematical discussion is based, are recognized to have one serious limitation, that is, they do not explain what causes or drives the electric and magnetic fields to propagate from one point to another.  In other words, the wave equations themselves shed little light on the mechanism of light propagation.  In the following, the mechanism of light propagation is first explained.

The physics behind light propagation can be understood from Faraday's law
\begin{equation}
\label{e8}
\nabla \times \vec{E}=-\frac{1}{c}\frac{\partial}{\partial t}\vec{B},    
\end{equation}
Amp\`{e}re's law
\begin{equation}
\label{e9}
\nabla \times \vec{B}=\frac{4\pi}{c}\vec{j}+\frac{1}{c}\frac{\partial}{\partial t}\vec{E},  
\end{equation}
and, in particular, from what these laws mean.  (See Refs. \cite{her07,dys91} for attempts to derive Maxwell equations from the continuity equation and Newton's second law, respectively.) As in Ref. \cite{jac75}, these laws are interpreted in the present work as follows.  While in Faraday's law $\vec{B}$ is the cause of $\vec{E}$, in Amp\`{e}re's law $\vec{B}$ is the effect of $\vec{E}$ and $\vec{j}$.  (See, for example, Refs. \cite{jef04,hil10,sav12,kin20}, for other interpretations.) This interpretation is adopted, because it can explain how light propagates.  A magnetic field creates, in its neighborhood, an electric field (Faraday's law), and the electric field then creates another magnetic field (Amp\`{e}re's law) further away from the current density.  Such a process of alternate creation of electric and magnetic fields is a never ending process, through which light propagates in vacuum.  This picture of light propagation resembles that of a mechanical wave.  In mechanics, the motion of a particle usually causes another particle next to it to move as a result of particle-particle interaction, and the motion of the latter particle subsequently causes its neighboring particle, the particle that is further away from the first one, to move too.  It is through such particle-particle interaction that motion is transferred from one particle to another, and a mechanical wave is formed \cite{guo13}.  (Also accounted for in Ref. \cite{guo13} is why in a macroscopic system composed of many particles a mechanical wave always travels with a finite speed.)

In the light of the preceding explanation of light propagation, it becomes convenient to understand physically why the emitted electric field must depend on the odd-order time derivatives of $\vec{j}$ and why the emitted magnetic field must depend instead on the even-order time derivatives of $\vec{j}$.  From Amp\`{e}re's law [see Eq. (\ref{e9})], it is evident that the current density cannot create an electric field directly, because it has to create a magnetic field first, and then, according to Faraday's law [see Eq. (\ref{e8})], it is the magnetic field that gives rise to an electric field, a field that depends on the first-order time derivative of $\vec{j}$.  The electric field subsequently becomes the source of another magnetic field, see Amp\`{e}re's law, and the latter magnetic field---a function of the second-order time derivative of $\vec{j}$---acts, as well, as the source of another new electric field (which depends on the third-order time derivative of $\vec{j}$), see Faraday's law.  It is through this process, time derivatives of $\vec{j}$ are built into the electric and magnetic fields that are emitted from $\vec{j}$ to the far-field region.  Light observation in the far-field region next requires that the current density should be differentiable in time infinitely, because the said alternate creation of electric and magnetic fields has to be maintained until the emitted electric field and the emitted magnetic field reach the far-field region.  Thus, the present hypotheses can also find support from physics.  In the following, the support from physics is made quantitative by deriving the electromagnetic wave equations in the present picture of light propagation, that is, the alternate creation of electric and magnetic fields will be shown to lead unambiguously to wave propagation of the emitted electric and magnetic fields. 

The first-order magnetic field $\vec{B}^{(1)}$ can only come from $\vec{j}$, that is, according to Amp\`{e}re's law,
\begin{equation}
\label{e10}
\nabla \times \vec{B}^{(1)}=\frac{4\pi}{c}\vec{j},
\end{equation}
because $\rho$ is excluded in this work and, thus, cannot create an electric field to be used in Eq. (\ref{e10}).  Following Faraday's law, $\vec{B}^{(1)}$ becomes the source of the first-order electric field $\vec{E}^{(1)}$: 
\begin{equation}
\label{e11}
\nabla \times \vec{E}^{(1)}=-\frac{1}{c}\frac{\partial}{\partial t}\vec{B}^{(1)}.
\end{equation}
Combine the preceding two equations to express $\vec{E}^{(1)}$ in terms of $\vec{j}$,
\begin{equation}
\label{e12}
\nabla ^{2} \vec{E}^{(1)}=\frac{4\pi}{c^{2}}\frac{\partial}{\partial t}\vec{j}.   
\end{equation}
Since $\vec{E}^{(1)}$ is the effect of $\vec{B}^{(1)}$, not some charger distribution, it is a non-Coulomb field satisfying $\nabla \cdot \vec{E}^{(1)}=0$.  Note also that Eq. (\ref{e12}) is not a wave equation, meaning that there is no time delay between $\vec{E}^{(1)}$ and $\vec{j}$.  If the first term on the right-hand side of Eq. (\ref{e7}) is denoted as $\vec{E}_{1}$, then it satisfies the same differential equation as that in Eq. (\ref{e12}).  Both $\vec{E}^{(1)}$ and $\vec{E}_{1}$ depend on the first-order time derivative of $\vec{j}$. 

Since $\vec{j}$ has every order of time derivative, the electric and magnetic fields $\vec{E}^{(1)}$ and $\vec{B}^{(1)}$ can never be the only fields at position $\vec{r}$.  From Amp\`{e}re's law, $\vec{E}^{(1)}$ must be responsible for another magnetic field denoted as the second-order field $\vec{B}^{(2)}$:
\begin{equation}
\label{e13}
\nabla \times \vec{B}^{(2)}=\frac{1}{c}\frac{\partial}{\partial t}\vec{E}^{(1)}.    
\end{equation}
Note that $\vec{j}$ is only responsible for $\vec{B}^{(1)}$, not $\vec{B}^{(2)}$.  Similarly, $\vec{B}^{(2)}$ in turn produces a different electric field denoted as the second-order electric field $\vec{E}^{(2)}$:
\begin{equation}
\label{e14}
\nabla \times \vec{E}^{(2)}=-\frac{1}{c}\frac{\partial}{\partial t}\vec{B}^{(2)}.    
\end{equation}
Again, since $\vec{E}^{(2)}$ is a non-Coulomb electric field, the preceding two equations lead to the following relation:
\begin{equation}
\label{e15}
\nabla ^{2} \vec{E}^{(2)}=\frac{1}{c^{2}}\frac{\partial ^{2}}{\partial t^{2}}\vec{E}^{(1)}. \end{equation}
Now that $\vec{E}^{(1)}$ depends on the first-order time derivative of $\vec{j}$, Eq. (\ref{e15}) shows that the electric field $\vec{E}^{(2)}$ must depend on the third-order time derivative of $\vec{j}$.  Again in Eq. (\ref{e7}), the second term on the right-hand side, if denoted as $\vec{E}_{2}$, is found to satisfy, with $\vec{E}_{1}$, the same equation as that in Eq. (\ref{e15}).

Following the same procedure to apply Amp\`{e}re's law and Faraday's law repeatedly, it turns out that at position $\vec{r}$ there is a series of electric fields, each of which depends on an odd-order time derivative of $\vec{j}$.  In particular, these electric fields are related to each other through relations similar to that in Eq. (\ref{e15}):
\begin{equation}
\label{e16}
\nabla ^{2}\vec{E}^{(n+1)}=\frac{1}{c^{2}}\frac{\partial ^{2}}{\partial t^{2}}\vec{E}^{(n)},    
\end{equation}
where $n=1, 2, \cdots$.  Add Eqs. (\ref{e12}), (\ref{e15}), and (\ref{e16}) to yield
\begin{eqnarray}
\label{e17}
\nabla ^{2}\Big (\vec{E}^{(1)}+\vec{E}^{(2)}+\cdots \Big )&=&\frac{4\pi}{c^{2}}\frac{\partial}{\partial t}\vec{j}+\frac{1}{c^{2}}\frac{\partial ^{2}}{\partial t^{2}}\Big (\vec{E}^{(1)}\nonumber\\
& &+\vec{E}^{(2)}+\cdots
\Big ).   
\end{eqnarray}
If $\vec{E}_{T}$ is defined as $\sum _{n=1}\vec{E}^{(n)}$, then it satisfies a wave equation no other than that in Eq. (\ref{e3}): 
\begin{equation}
\label{e18}
\nabla ^{2}\vec{E}_{T}-\frac{1}{c^{2}}\frac{\partial ^{2}}{\partial t^{2}}\vec{E}_{T}=\frac{4\pi}{c^{2}}\frac{\partial }{\partial t}\vec{j}.    
\end{equation}

Since it satisfies the wave equation (\ref{e18}), where $\vec{j}$ is the source, the net field $\vec{E}_{T}$ must be the electric field that is emitted from $\vec{j}$ and observed in vacuum.  Experimentally, the net field $\vec{E}_{T}$ always has a finite magnitude, meaning that the series $\sum _{n}\vec{E}^{(n)}$ must be convergent.  Still from the derivation of Eq. (\ref{e18}) and the definition of $\vec{E}_{T}$, those terms on the right-hand side of Eq. (\ref{e7}) can now be explained physically to correspond to the individual electric fields.  

Note that, as Eqs. (\ref{e12}) and (\ref{e15}) show, neither of the individual electric fields $\vec{E}^{(n)}$ satisfies a wave equation.  Practically, these individual electric fields can be interpreted as instantaneous fields, fields that can reach any position from $\vec{j}$ without any time delay.  This interpretation nevertheless does not mean that the principle of causality is violated in light propagation, because, experimentally, these individual fields are not observed.  What is observed is the sum of the individual fields $\vec{E}_{T}$, and $\vec{E}_{T}$ itself satisfies a wave equation and thus is a wave propagating with speed $c$.  Causality is meaningful only in an observable process.  In classical electrodynamics, instantaneous quantities are not uncommon.  One example is the scalar potential in the Coulomb gauge \cite{jac75,gar88}.  See also Ref. \cite{her05}. 

Similar analysis shows that at position $\vec{r}$, there is also a series of magnetic fields $\vec{B}^{(n)}$, and the net magnetic field $\vec{B}_{T}$ defined as $\sum _{n=1}\vec{B}^{(n)}$ satisfies a wave equation identical to that in Eq. (\ref{e2}):
\begin{equation}
\label{e19}
\nabla ^{2}\vec{B}_{T}-\frac{1}{c^{2}}\frac{\partial ^{2}}{\partial t^{2}}\vec{B}_{T}=-\frac{4\pi}{c}\nabla \times \vec{j}.    
\end{equation}
Thus, $\vec{B}_{T}$ must be the magnetic field emitted from $\vec{j}$ and observed in vacuum.  Unlike $\vec{E}^{(n)}$, however, the individual magnetic fields $\vec{B}^{(n)}$ each depend on an even-order time derivative of $\vec{j}$.  

Like many other radiative processes in nature, light emission from $\vec{j}$ is observed to satisfy the principle of causality.  Guided by this observation, the solutions to the wave equations in Eqs. (\ref{e2}) and (\ref{e3}) are conventionally chosen to be retarded \cite{pan62,jac75,zan12}:
 \begin{equation}
\label{e20}
\vec{E}_{ret}(\vec{r},t)=-\frac{1}{c^{2} }\int _{V_{2}} \frac{1}{\vert \vec{r}-\vec{r}_{1}\vert }\frac{\partial}{\partial t}\vec{j}\big (\vec{r}_{1},t-\vert \vec{r}-\vec{r}_{1}\vert c^{-1}\big )d\vec{r}_{1},    
\end{equation}
and
\begin{equation}
\label{e21}
\vec{B}_{ret}(\vec{r},t)=-\frac{1}{c}\int _{V_{2}}\frac{1}{\vert \vec{r}-\vec{r}_{1}\vert }\nabla _{1} \times \vec{j}\big (\vec{r}_{1},t-\vert \vec{r}-\vec{r}_{1}\vert c^{-1}\big )d\vec{r}_{1},    
\end{equation}
where $\nabla _{1}$ operates on the leftmost $\vec{r}_{1}$ in $\vec{j}(\vec{r}_{1},t-\vert \vec{r}-\vec{r}_{1}\vert c^{-1})$.  See Ref. \cite{her07} for mathematical operations on retarded quantities.  Since $\vec{j}$ is infinitely differentiable in time, it is always valid to expand $\vec{j}(\vec{r}_{1},t-\vert \vec{r}-\vec{r}_{1}\vert c^{-1})$ in the retarded solutions into, for example, a Taylor series around $t$.  Such expansion immediately shows that, unlike $\vec{E}_{T}$ or $\vec{B}_{T}$, the retarded solutions depend on not only the odd-order time derivatives of $\vec{j}$ but also the even-order time derivatives, a property that can be explained by neither mathematics nor physics.  For this reason, it is fair to conclude that although they are well known, the retarded solution $\vec{E}_{ret}$ can never be the emitted electric field, and the retarded solution $\vec{B}_{ret}$ can never be the emitted magnetic field.  
 
To summarize, it is found, mathematically and physically, that in order for a time-dependent current density to be a source of electromagnetic radiation, the current density has to be differentiable in time infinitely.  Each order of the time derivative corresponds to an individual electric or magnetic field.  The sum of the individual fields becomes the usual electromagnetic radiation that is emitted from the current density and propagates in vacuum with speed $c$.  Other properties of the radiation are also discussed.

\section*{Disclosure statement}
The author declares no conflicts of interest.

\end{document}